\documentclass[12pt]{article}
\usepackage{graphicx}
\begin{document}

\title{Electrical Conductance of Molecular Wires}
\author{Eldon Emberly\footnote{email: eemberly@sfu.ca  copyright IOP 
publishing 1999} and George Kirczenow,\\ Department of Physics,\\ 
Simon Fraser University, \\Burnaby, B.C., Canada\\ V5A
1S6}

\date{\today}

\maketitle
\begin{abstract}
Molecular wires (MW) are the fundamental building blocks for
molecular electronic devices.  They consist of a molecular unit
connected to two continuum reservoirs of electrons (usually
metallic leads).  We rely on Landauer theory as
the basis for studying the conductance properties of MW
systems.  This relates the lead to lead current to the
transmission probability for an electron to scatter through the
molecule.  Two different methods have been developed for the
study of this scattering.  One is based on a solution of the
Lippmann-Schwinger equation and the other solves for the {\bf
t} matrix using Schroedinger's equation.  We use our
methodology to study two problems of current interest.  The
first MW system consists of 1,4 benzene-dithiolate (BDT) bonded
to two gold nanocontacts.  Our calculations show that the
conductance is sensitive to the chemical bonding between the
molecule and the leads.  The second system we study highlights
the interesting phenomenon of antiresonances in MW.  We derive
an analytic formula predicting at what
energies antiresonances should occur in the transmission
spectra of MW. A numerical
calculation for a MW consisting of
filter molecules attached to an active molecule shows the
existence of an antiresonance at the energy predicted by our
formula.
\end{abstract}
\section{Introduction}
A molecular wire (MW)in its simplest definition consists of a
molecule connected between two reservoirs of electrons.  The
molecular orbitals of the molecule when they couple to the
leads provide favourable pathways for electrons. Such a system
was suggested in the early '70's by Aviram and Ratner to have
the ability to rectify current\cite{Avi74}.  Experimental
research on MW has increased over the past few years looking
into the possibility of rectification and other
phenomena\cite{Zhou97,Reed97,Frank98,Tans97,Andres96,Mirkin96,
Bumm96, Datta97_2}. There has also been an increase in the
theoretical modeling of MW
systems\cite{Ember98_1,Mago97,Datta97_1,Samant96,Kemp96,
Mujic96}. For a comprehensive overreview of the current status
of the molecular electronics field see \cite{conf}.

Theoretical studies of the electronic conductance of a MW bring
together different methods from chemistry and physics. Quantum
chemistry is used to model the energetics of the molecule.  It
is also incorporated into the study of the coupling between the
molecule and the metallic reservoirs. Once these issues have
been addressed it is possible to proceed to the electron
transport problem.  Currently, Landauer
theory\cite{Lan57,Datta95} is used which relates the
conductance to the electron transmission probability.

A molecule of current experimental interest as a MW is 1,4
benzene-dithiolate (BDT)\cite{Reed97}. It consists of a benzene
molecule with two sulfur atoms attached, one on either end of
the benzene ring.  The sulfurs bond effectively to the gold
nanocontacts and the conjugated $\pi$ ring provides delocalized
electrons which are beneficial for transport.  Two major
unknowns of the experimental system are the
geometry of the gold contacts and the nature of the bond
between the molecule and these contacts.  This paper attempts
to highlight these important issues by showing how the
differential conductance varies with bond strength.

For mesoscopic systems with discrete energy levels (such as MW)
connected to continuum reservoirs, the transmission probability
displays resonance peaks.  Another potentially important
transport phenomenon that has been predicted is the appearance
of antiresonances\cite{Kemp96,Ember98_2}. These occur when the
transmission probability is zero and correspond to the incident
electrons being perfectly reflected by the molecule.  We derive
a simple condition controlling where the antiresonances occur
in the transmission spectrum.  We apply our formula to the case
of a MW consisting of an ``active'' molecular segment connected
to two metal contacts by a pair of finite $\pi$ conjugated
chains. In this calculation we show how an antiresonance can be
generated near the Fermi energy of the metallic leads.  The
antiresonance is characterized by a drop in conductance.
We find that for this calculation our analytic theory of
antiresonances has predictive power.

In Sec. II, we describe the scattering and transport theory
used in this work.  The first subsection deals with Landauer
theory.  The second subsection outlines a method for evaluating
the transmission probability for MW systems.  The last
subsection outlines our derivation of the antiresonance
condition.  In Sec. III, we present some calculations on BDT
for different lead geometries and different binding strengths.
Sec. IV describes a numerical calculation for a system
displaying antiresonances which are predicted by the formula in
Sec. II. Finally we conclude with some remarks in Sec. V.
\section{Transport Theory}
\subsection{Landauer Formula}
\begin{figure}[ht]
\begin{center}
\label{fig1}
\includegraphics[bb= 50 200 550 590
,width=0.75\textwidth,clip]{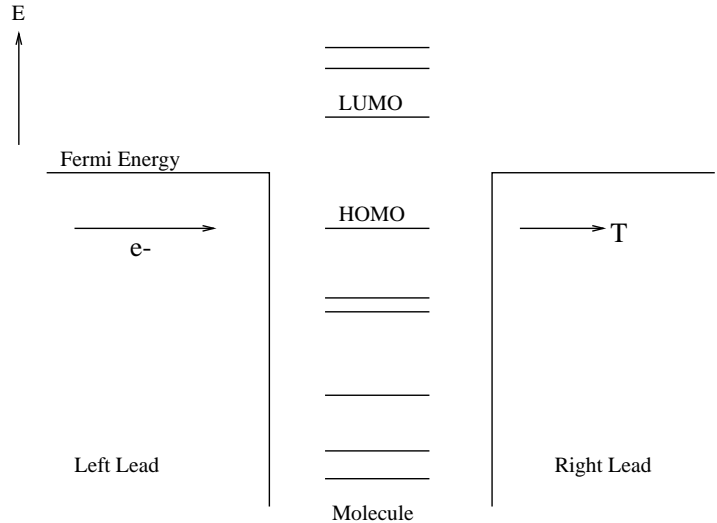}
\caption{A schematic energy diagram for a molecular wire.  The
leads couple to the energy levels of the molecule and electrons
are incident from the left lead at its Fermi energy.  They
scatter through the molecule and have a transmission
probability, $T$ to enter the right lead.}
\end{center}
\end{figure}

We consider the transport of electrons through a molecular
system by modeling it as a one electron elastic scattering
problem.  The molecule acts as a defect between two metallic
reservoirs of electrons.  An electron incident from the source
lead with an energy $E$, has a transmission probability $T(E)$
to scatter through the molecule into the drain lead.  A
schematic of the model system is shown in Fig. \ref{fig1}.  By
determining the transmission probability for a range of
energies around the Fermi energy $\epsilon_{F}$ of the source
lead, the finite temperature, finite voltage,
Landauer formula can be used to calculate the
transmitted current $I$ as a function of the bias voltage, $V$,
applied between the source (left lead) and drain (right lead)
\begin{equation}
I(V) = \frac{2e}{h} \int_{-\infty}^{\infty}
dE\:T(E)\left( \frac{1}{\exp[(E-\mu_{s})/kT] + 1} -
\frac{1}{\exp[(E-\mu_{d})/kT]+1} \right)
\label{eq:Landauer}
\end{equation}
The two electro-chemical potentials $\mu_{s}$ and $\mu_{d}$,
refer to the source and drain, respectively.  They are defined
to be, $\mu_{s} = \epsilon_{F} + eV/2$ and $\mu_{d} =
\epsilon_{F} - eV/2$.  The differential conductance is then
given by the derivative of the current with respect to voltage.

\subsection{{\bf t} Matrix Method}
We find the transmission probability $T(E)$ used in the Landauer
formula, Eq.(\ref{eq:Landauer}), by solving the Schroedinger equation
directly for the scattered wavefunction of the electron.  The electron
is initially propagating in a Bloch wave in one of the modes of the
source lead.  The molecule will reflect some of this wave back into
the various modes of the source lead.  The molecule is represented by
a discrete set of molecular orbitals (MO's) through which the electron
can tunnel.  Hence, some of the wave will be transmitted through the
molecule and into the modes of the drain lead. By finding the
scattered wavefunction it is then possible to determine how much was
transmitted, yielding $T(E)$.

We start with Schroedinger's equation, $H|\Psi^\alpha\rangle =
E|\Psi^\alpha\rangle$, where $H$ is the Hamiltonian for the
entire MW system consisting of the leads and the molecule.
$|\Psi^\alpha\rangle$ is the wavefunction of the electron
propagating initially in the $\alpha^{th}$ mode of the left
lead with energy $E$.  It is expressed in terms of the
transmission and reflection coefficients, $t_{\alpha,\alpha'}$
and $r_{\alpha,\alpha'}$ and has different forms on the left
lead (L), molecule (M), and right lead (R).  The total
wavefunction is a sum of these three, $|\Psi^{\alpha} \rangle =
|\Psi^{\alpha}_{L}\rangle + |\Psi^{\alpha}_{M}\rangle +
|\Psi^{\alpha}_{R}\rangle$, where
\begin{eqnarray}
|\Psi^{\alpha}_{L}\rangle &=& |\Phi_{+}^{\alpha}\rangle +
\sum_{\alpha'} r_{\alpha',\alpha} |\Phi_{-}^{\alpha'}\rangle \\
|\Psi^{\alpha}_{M}\rangle &=& \sum_j c_j |\phi_j\rangle \\
|\Psi^{\alpha}_{R}\rangle &=&  \sum_{\alpha'} t_{\alpha,\alpha'}
|\Phi_{+}^{\alpha'}\rangle
\end{eqnarray}
In the above, $|\Phi_{\pm}^{\alpha}\rangle$ are
forward/backward propagating Bloch waves in the $\alpha^{th}$
mode, and $|\phi_j\rangle$ is the $j^{th}$ MO on the molecule.

We then consider our Hamiltonian in the tight-binding (or
H\"{u}ckel) approximation and use Schroedinger's equation to
arrive at a system of linear equations in the unknown
quantities, $r_{\alpha',\alpha}$, $c_j$ and
$t_{\alpha',\alpha}$.  We solve numerically for the reflection
and transmission coefficients for each rightward propagating
mode $\alpha$ at energy $E$ in the left lead.  The total
transmission is then given by
\begin{equation}
T(E) = \sum_{{\alpha \epsilon L}} \sum_{{\alpha'
\epsilon R}} \frac{v^{\alpha'}}{v^{\alpha}}
|t_{\alpha',\alpha}|^{2}
\end{equation}
where $v^{\alpha}$ is the velocity of the electron in the
$\alpha^{th}$ rightward propagating mode.  The sum over
$\alpha'$ in the expression for $T$ is over the rightward
propagating modes at energy $E$ in the right lead.  With $T(E)$
determined the current and differential conductance can then be
calculated using Eq. (\ref{eq:Landauer}).
\subsection{Antiresonance Condition}
The above section outlined a method for numerically evaluating
the transmission probability for a MW system.  We now focus on
an interesting physical feature of the transmission
probability, namely the occurrence of antiresonances.  These
are characterized by the perfect reflection of electrons
incident on the molecule at a certain energy $E$.  The basis of
our analysis in this section will be the Lippmann-Schwinger
(LS) equation applied to the scattering problem in a highly
idealized model. It will lead to an analytic formula for $T(E)$.

Before we proceed further we wish to highlight a problem that
arises in most quantum chemistry applications.  In studying
quantum systems, such as MW, which are composed out of atomic
building blocks, it is customary to solve the problem by
expressing the electron's wavefunction in terms of the atomic
orbitals on the various atoms.  A problem that arises is that
these orbitals are not usually orthogonal to each
other. Including this nonorthogonality complicates the solution
and so it is often neglected.  In the analysis that follows we
have utilized a transformation which removes the
nonorthogonality and allows for a straightforward
solution\cite{Ember98_2}. This transformation leads to an
energy dependent Hamiltonian which will play an important role
in determining the presence of antiresonances.

The LS equation we consider has the form
\begin{equation}
|\Psi ' \rangle = |\Phi ' \rangle + G'(E) W^E |\Psi ' \rangle .
\label{eq:LS}
\end{equation}
Here $|\Psi'\rangle$ is the scattered electron wavefunction of the
transformed Hamiltonian $H^E$.  $W^E$ couples the molecule to the
adjacent lead sites.  $|\Phi' \rangle$ is the initial electron state
which is a propagating Bloch wave that is confined to the left lead.
$G'(E)=(E - H_0^E)^{-1}$ is the Green's function of the decoupled
system.

We consider the LS equation in the tight binding approximation
and solve for $|\Psi '\rangle$.  This gives us the value for
$\Psi_1$ which is the value of the wavefunction on the first
atomic site on the right lead.  The transmission probability,
$T$ is simply $|\Psi_1|^2$. The result is
\begin{equation}
\Psi_{1} = \frac{P \Phi'_{-1}}{[(1-Q)(1-R)-PS]} \label{eq:psi1}
\end{equation}
where
\begin{eqnarray*}
P &=& G'_{1,1} \sum_j W^E_{1,j} G'_j W^E_{j,-1} \\
Q &=& G'_{1,1} \sum_j (W^E_{1,j})^2 G'_j \\
R &=& G'_{1,1} \sum_j (W^E_{-1,j})^2 G'_j \\
S &=& G'_{1,1} \sum_j W^E_{-1,j} G'_j W^E_{j,1}
\end{eqnarray*}
The sum over $j$ is over only the MO's.  In the above,
$W^E_{1,j} = H_{1,j} - E S_{1,j}$ is the energy-dependent
hopping element of $H^E$ between the first lead site and the
$j^{th}$ MO in terms of the hopping element of the original
Hamiltonian $H$ and the overlap $S$ in the non-orthogonal
basis. The Green's function on the molecule is expanded in
terms of its molecular eigenstates and this gives $G'_j =
1/(E-\epsilon_j)$ for the $j^{th}$ MO with energy
$\epsilon_j$. $G'_{1,1}$ is the diagonal matrix element of the
Green's function $G'(E)$ at the end site of the isolated lead.

Antiresonances of the MW occur where the
transmission $T$ is equal to zero.   These occur at Fermi
energies $E$ that are the roots of Eq. (\ref{eq:psi1}), namely
\begin{equation}
\sum_j \frac{(H_{1,j} - E S_{1,j})(H_{j,-1} - E S_{j,-1})}{E -
\epsilon_j} = 0.
\label{eq:anti}
\end{equation}

Antiresonances can arise due to an interference between
molecular states that may differ in energy, as is seen directly
from Eq. (\ref{eq:anti}). An electron hops from the left lead
site adjacent to the molecule onto each of the MO's with a
weight $W^E_{j,-1}$. It then propagates through each of the
different orbitals and hops onto the right lead with a
weight $W^E_{1,j} \,.$ These processes interfere with each
other and where they cancel (\ref{eq:anti}) is satisfied and an
antiresonance occurs.

Antiresonances can also arise solely from the nonorthogonality
of atomic orbitals. This occurs when
only a single MO $k$ couples appreciably to the leads.  Eq.
(\ref{eq:anti}) then becomes $(H_{k,-1} - E S_{k,-1})(H_{1,k} -
E S_{1,k}) = 0$.  This equation gives two antiresonances. The
non-orthogonality of two orbitals can actually stop electron
hopping between these orbitals, which blocks electron transport
and creates an antiresonance. This is a counterintuitive effect
since one would normally expect orbital overlap to aid electron
transfer between the orbitals rather than hinder it. Thus the
inclusion of orbital overlap leads to non-trivial physical
consequences in the tight binding treatment of MW.
\section{Conductance of BDT}
\begin{figure}[ht]
\begin{center}
\includegraphics[bb= 150 270 480 675
,width=0.5\textwidth,clip]{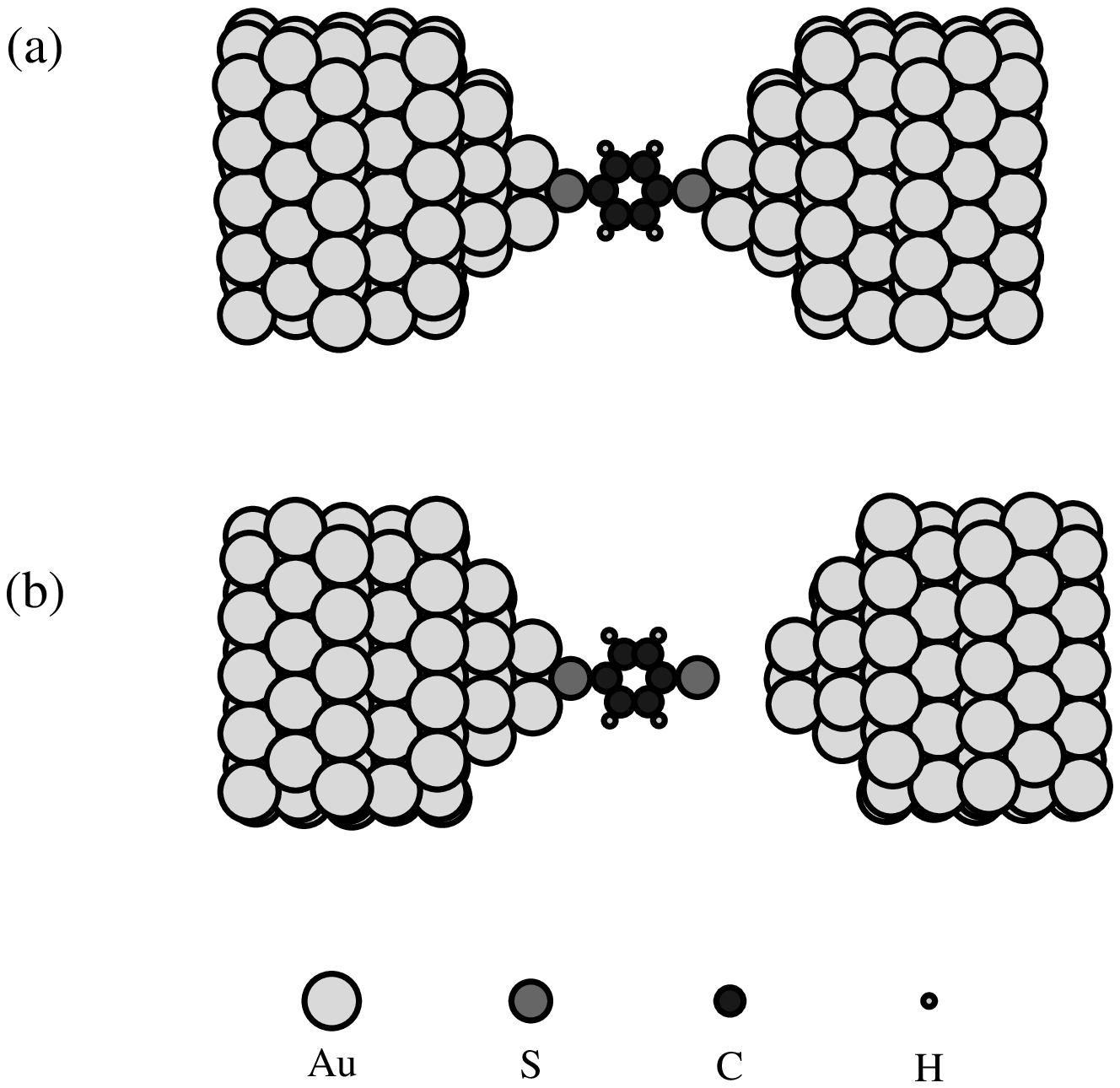}
\caption{Atomistic diagram of gold (111) leads bonded to a BDT
molecule.  (a) Case of strong bonding to both leads. (b) Case
of weak bonding to one of the leads.}
\label{fig2}
\end{center}
\end{figure}

For our first application, we apply our {\bf t} matrix
formalism to study the conductance of 1,4 benzene-dithiolate, a
MW of recent experimental interest\cite{Reed97}. The molecule
consists of a benzene molecule with the hydrogen atoms at the 1
and 4 positions replaced with sulphur atoms.  The sulphur atoms
act like alligator clips when they bond to the gold leads.  We
calculate the conductance of the MW geometries shown in
Fig. \ref{fig2}.  The gold leads are oriented in the
(111) direction. Attached to the molecule are gold clusters
which form the tips for the lead.  Experimentally it has been
found that sulphur atoms preferentially bind over the hollow
sites formed on gold surfaces and so for our simulation the BDT
molecule is bonded over the hollow site on each tip.

The strength of the bond between the molecule and the gold
surface plays an important role in determining the transmission
characteristics of the BDT MW.  Isolated BDT has a discrete set
of MO's with the highest occupied molecular orbital (HOMO)
calculated to be around -10.5 eV and the lowest unoccupied
molecular orbital (LUMO) found to be around -8.2 eV.  These
levels when bonded with the leads become part of the continuum
of energy states that exist within the metallic reservoirs.
For strong bonding they can become significantly altered as
their chemical nature becomes mixed with the surface states of
the gold tips.  For weaker bonding the MO's retain the
character of the isolated molecule.

\begin{figure}[ht]
\begin{center}
\includegraphics[bb= 10 50 550 750
,width=0.65\textwidth,clip]{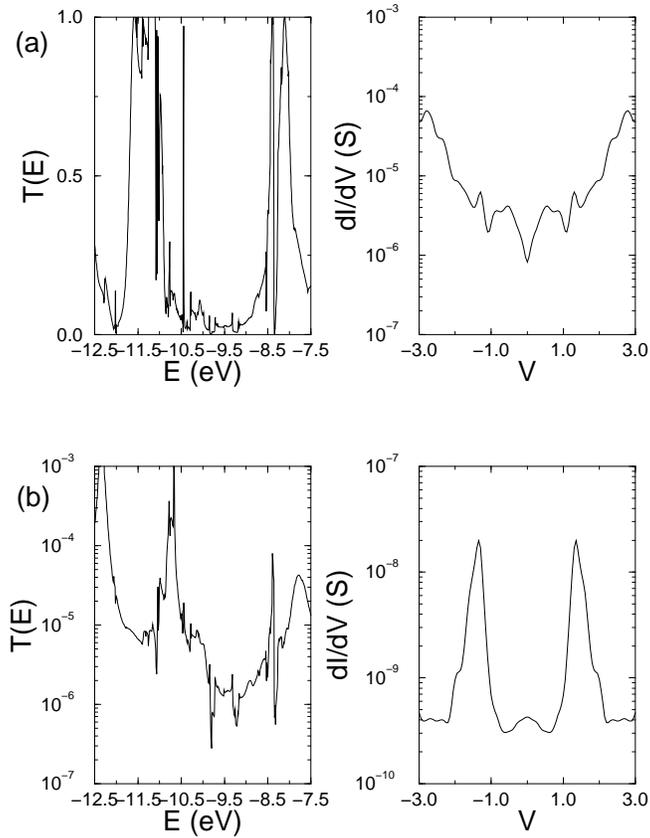}
\caption{(a)Transmission and differential conductance for BDT
bonded to gold (111) leads with strong coupling. (b) weak
coupling case.}
\label{fig3}
\end{center}
\end{figure}

We consider strong binding to the (111) leads first.  The
transmission diagram is shown in Fig. \ref{fig3}a.  There is
strong transmission in the energy regions where the gold tip
states have mixed with the molecular states.  This occurs most
prominently around -11.5 eV, where there exist resonances that
can be connected with the HOMO states of the molecule. The HOMO
and states around it in the isolated molecule have been mixed
and lowered in energy due to the bonding to the lead.  The
other region of significant transmission is at around -8 eV,
which is due to states connected with the LUMO of the BDT. The
region in between has resonances that arise from those states
that are complex admixtures of gold tip and molecule levels.
The differential conductance was calculated with a Fermi level
chosen at -10 eV which lies in the HOMO-LUMO gap.  The molecule
seems to be very conductive when attached strongly to the (111)
oriented wide leads.

The magnitude of the conductance found in the above calculation
exceeds that found experimentally\cite{Reed97}. We now
consider the case of weak binding to see what effect it will
have on the transmission.  One lead is allowed to bond strongly
to the molecule, while the other lead is pulled away from the
molecule so as to be at non-bonding distances.  The
transmission and conductance diagrams for the weak bonding case
are shown in Fig. \ref{fig3}b.  It can be seen that the
transmission has gone down significantly in the region between
-10.5 and -8 eV. There are also now strong resonances that
correspond almost exactly with the isolated molecular levels.
The weaker coupling of the molecule to the contacts has lowered
the transmission and the conductance.  Because there are order
of magnitude fluctuations in $T(E)$ the conductance
is now quite sensitive to the selection of Fermi energy.
Different conductance curves can be obtained if the Fermi
energy is chosen to occur on or off resonance. 

\section{Antiresonance Calculations}
We now look at a MW system that displays antiresonances that
are predicted by the formula derived in Sec. II.  Eq. (\ref{eq:anti})
 was derived for an idealized MW system where there was
only one incident electron mode on the molecule.  Thus in
designing the MW system for the numerical study we tried to
approximate the ideal system as closely as possible.  The MW
system we suggest consists of left and right $\pi$ conjugated
chain molecules attached to what we will call the ``active''
molecule.  The purpose of these conjugated chains is to act as
filters to the many modes that will be incident from the
metallic leads.  For appropriate energies they will restrict
the propagating electron mode to be only $\pi$ like.  This
$\pi$ backbone will only interact with the $\pi$ orbitals of the
molecule if they are bonded in an appropriated fashion.

\begin{figure}[ht]
\begin{center}
\includegraphics[bb= 35 237 600 470
,width=0.5\textwidth,clip]{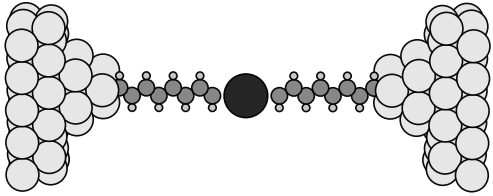}
\caption{Atomistic diagram of leads attached to filter
molecules and active molecule.  The filter molecules consist of
C$_8$H$_8$ chains.  The active molecule contains 2 $\pi$ levels
which interact with the filter molecules.  The leads are (111)
gold.}
\label{fig4}
\end{center}
\end{figure}

An atomic diagram of our system is shown in Fig. \ref{fig4}.
The leads are oriented in the (111) direction.  The chain
molecules are bonded to clusters of gold atoms that form the
tips of the leads.  The carbon atom nearest to the gold tip
binds over the hollow site of the tip.
The chains each have eight CH groups, which for the
energies considered will only admit a $\pi$ like state to
propagate along them. This gives rise to the filtering process
mentioned above.  The active molecule is chosen to have two
$\sigma$ states and two $\pi$ states. The chain's $\pi$ like
orbitals will only couple to the two $\pi$ states of the active
molecule.

\begin{figure}[ht]
\begin{center}
\includegraphics[bb= 75 85 510 780
,width=0.5\textwidth,clip]{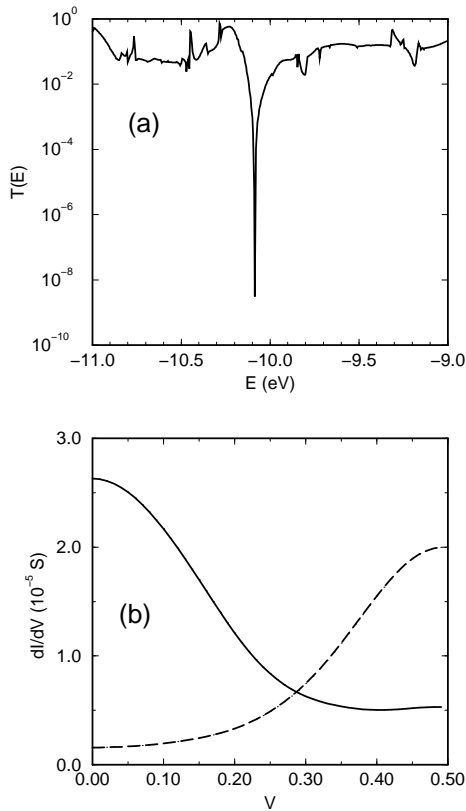}
\caption{(a) Transmission diagram for antiresonance molecule
displaying a prominent antiresonance at -10.1 eV. (b)
Differential conductance for $\epsilon_F$ =-10.2 eV (solid
line) and $\epsilon_F$ = =10.0 eV (dashed curve).}
\label{fig5}
\end{center}
\end{figure}

The Fermi energy for our gold leads is around -10 eV which lies
within the $\pi$ band.  We would like an antiresonance to occur
somewhere near this energy.  The parameters entering
Eq. \ref{eq:anti} (i.e. the molecular orbital energies and
their overlaps with the chains) are chosen so that one of the
roots of the equation yields a value near -10 eV.  The
numerically calculated electron transmission probability for
this model is shown in Fig. \ref{fig5}a; an antiresonance is
seen in the plot at the predicted energy of -10.08 eV.  The
differential conductance at room temperature was calculated
using Eq. (\ref{eq:Landauer}).  Two different conductance
calculations are shown in Fig. \ref{fig5}b.  The solid curve
corresponds to a choice of Fermi energy of -10.2 eV.  Because
it lies to the left of the antiresonance in a region of strong
transmission, the conductance is strong at 0 V.  It then drops
at around 0.2 V when the antiresonance is crossed.  The dashed
curve was calculated using a Fermi energy of -10.0 eV.  It
starts in a region of lower transmission and thus the
antiresonance suppresses the increase in current.  After 0.2 V
the large transmission to the left of the antiresonance is
sampled and the current rises sharply.  So in both cases the
antiresonance has lowered the conductance. It is conceivable to
think of utilizing more antiresonances in a narrow energy range
to create a more pronounced conductance drop.

\section{Conclusions}
We have presented a theoretical study of the electrical
conductance properties of molecular wires in the context of
one-electron theory.  A numerical method has been developed for
the study of transport in molecular wires which solves for the
transmission coefficients using Schroedinger's equation.  It
allows for the study of multimode leads attached to a molecule.
We then proceeded to present a simple analytically solvable
model which highlighted the interesting phenomena of
antiresonances.  A formula was derived that predicts Fermi
energies for which a molecule with a given set of molecular
energy levels should display antiresonances.  It predicts two
mechanisms by which antiresonances arise: one due to
interference between the molecular orbitals and the other due
to a cancellation of the effective hopping parameter.

As an application of our numerical method studied the
conductance of BDT.  We examined the role of coupling by
considering both the strong and weak regimes.  For strong
coupling it was found that the MW has regions of strong
transmission.  These regions occur at energies which differ
from the isolated molecule's energy levels because of state
hybridization with the surface states of the gold tips.  The
conductance found for the strong coupling case was orders of
magnitude greater than that found experimentally, although
qualitatively it shared common features.  In the weak coupling
study the transmission displayed resonances at energies
corresponding to those of the isolated molecule.  The magnitude
was also significantly down.  This resulted in a conductance
curve that was of the magnitude found in the experiment.
Future work will need to focus on the electrostatic problem of
the molecule within an applied electric field and the
consequences of this in the context of Landauer theory,
many-electron and possible polaronic effects within the MW.

Using our analytic formula for antiresonances, we were able to
predict the occurrence of an antiresonance within a more
sophisticated numerical study utilizing a molecule attached
bridging a metal wire break junction. The model should also be
of interest for future MW work since it introduces the idea of
``filter'' molecules.  Our numerical studies showed that $\pi$
conjugated chain molecules act as effective mode filters to
electrons incident from the metallic leads.  The filter chains
in our model reduced the number of propagating modes down to
one, which was then coupled to the ``active'' molecule.  Our
formula was derived on the assumption of only a single
propagating mode, and for the ``active'' molecule considered it
was able to successfully predict the energy at which the
antiresonance occurred.  The antiresonance was charaterized by
a drop in the differential conductance.

%


\end{document}